\documentclass[10pt,twocolumn,twoside]{IEEEtran}
\usepackage{cite}
\usepackage{amsmath,amssymb,amsfonts}
\usepackage{textcomp}

\usepackage{ragged2e}
\usepackage{etoolbox}
\usepackage{lipsum}
\usepackage[pdftex]{graphicx}

\usepackage[english]{babel}
\usepackage{tikz}
\usepackage[T1]{fontenc}
\usepackage[utf8]{inputenc}
\usepackage{ae}\usepackage{latexsym}
\usepackage{graphicx,epsfig,psfrag}
\usepackage{amsmath}
\usepackage{rotate}
\usepackage{amsfonts}
\usepackage{bbold}
\usepackage{algorithm}
\usepackage[noend]{algpseudocode}

\newtheorem{definition}{Definition}

\newtheorem{example}{Example}

\newtheorem{theorem}{Theorem}
\newtheorem{lemma}{Lemma}

\newcommand {\be}{\begin{equation}}
\newcommand {\ee}{\end{equation}}
\def\qed{\hfill \vrule height 6pt width 6pt depth 0pt
          \smallskip}       

\usepackage{subfigure}

\usepackage{tikz}
\usepgflibrary {shadings} 
\usetikzlibrary{3d,shadings}

\usepackage{accents}
\newcommand{\ubar}[1]{\underaccent{\bar}{#1}}

\begin{document}
%

\title{{Characterization of Invariance, Periodic Solutions and Optimization of Dynamic Financial Networks}}

\author{Leonardo~Stella~\IEEEmembership{Member,~IEEE}, Dario~Bauso, Franco~Blanchini and Patrizio~Colaneri~\IEEEmembership{Fellow,~IEEE} 
%
\thanks{L. Stella is with the School of Computer Science, College of Engineering and Physical Sciences, University of Birmingham, Birmingham, B15 2TT, United Kingdom, e-mail: (l.stella@bham.ac.uk).}
\thanks{D. Bauso is with the Jan C. Willems Center for Systems and Control, ENTEG, Faculty of Science and Engineering, University of Groningen, The Netherlands, and with the Dipartimento di Ingegneria, University of Palermo, Italy, e-mail: (d.bauso@rug.nl).}%
\thanks{F. Blanchini is with Dipartimento di Matematica e Informatica, Universit\`a degli Studi di Udine, Italy, e-mail: (franco.blanchini@uniud.it).}%
\thanks{P. Colaneri is with Dipartimento di Elettronica, Informazione e Bioingeneria, and IEIIT-CNR, Politecnico di Milano, 20133 Milano, Italy, e-mail: (patrizio.colaneri@polimi.it).}%
}

\maketitle


\begin{abstract}  
Cascading failures, such as bankruptcies and defaults, pose a serious threat for the resilience of the global financial system. Indeed, because of the complex investment and cross-holding relations within the system, failures can occur as a result of the propagation of a financial collapse from one organization to another. While this problem has been studied in depth from a static angle, namely, when the system is at an equilibrium, we take a different perspective and study the corresponding dynamical system. The contribution of this paper is threefold. First, we carry out a systematic analysis of the regions of attraction and invariance of the system orthants, defined by the positive and negative values of the organizations' equity. Second, we investigate periodic solutions and show through a counterexample that there could exist periodic solutions of period greater than $2$. Finally, we study the problem of finding the smallest cash injection that would bring the system to the maximal invariant region of the positive orthant.

\end{abstract}

\begin{IEEEkeywords}
Systemic risk; Financial network; Financial contagion; Stability analysis; Invariance. 
\end{IEEEkeywords}
\IEEEpeerreviewmaketitle

\section{Introduction}
\IEEEPARstart{I}{n} the recent years, several episodes have exposed the instability and vulnerability of the global financial system, including, for instance, the collapses of Silicon Valley Bank and Credit Suisse (CS), or the critical situation of Greece and Spain in 2012 to the brink of default \cite{Elliott_2014}. Motivated by these episodes, in this paper we investigate the propagation of failures in financial systems.

The global financial market is a complex system, where a range of key players hold each other's shares, debts and obligations in different capacity. In the rest of this paper, we use the term \emph{organization} in a broad sense in a similar manner as it is used in the related literature to indicate these key players, including entities such as governments, banks, firms, companies, etc., or individuals, e.g., private citizens. Due to this large number of financial interdependencies, when an organization fails, this failure can produce cascading effects which can, as a result, lead to more defaults and bankruptcies. These interdependencies form a network of mutual and unilateral links among the organizations, which act as conduit for more cascading defaults.

In this paper, we investigate the role of cascading failures in the global financial system and the propagation of failures to organizations in the network. In our previous work \cite{Stella_2024}, we proposed a discrete-time dynamical system that describes the evolution of the equity values of a set of organizations and study the stability of its equilibrium points. Here, we briefly recall the main results and extend the study to invariance and the design of the minimal cash injection to mitigate the impact of financial contagion. In the model, it is worth highlighting three relevant aspects. First, the financial interdependencies among organizations are captured by the matrix of cross-holdings. This matrix represents the shares or other liabilities held by other organizations. Second, each organization can own external assets, which is captured by a network where each link connects the organizations to the market price of the owned assets. Finally, an endogenous nonlinearity is given by the cost of failure. When the equity value of an organization falls below a certain threshold, the organization has to pay this charge, which is responsible, together with the other interdependencies, for the propagation of failures.

\textit{Related Works}. In the pioneering work by Eisenberg and Noe in 2001 \cite{Eisenberg_2001}, the authors propose a framework that captures how shocks in the financial system propagate through a network of organizations in inter-bank lending. The contagion develops instantly at individual nodes and affects other nodes in the network, leading to new equilibrium points representing the agreed mutual payments.

In the following years, a substantial body of work have analyzed and generalized this framework, e.g., through the investigation of the impact of the network structure on the propagation of the shocks and their magnitude, with a focus on sparsity and asymmetry of the network \cite{Carvalho_2008, Acemoglu_2012}. Their model can accommodate a variety of settings. For instance, in production networks the model represents the input-output relationships and determines the output equilibrium \cite{Acemoglu_2012}. Another example in financial systems, their model can be used to calculate the clearing loan repayments~\cite{Eisenberg_2001}. Later, the preliminary work by Eisenberg and Noe was extended in several directions. The work by Elsinger \cite{Elsinger_2009} and then followed by Elliott \emph{et al.} \cite{Elliott_2014}, Rogers and Veraart \cite{Rogers_2013}, and Glasserman and Young \cite{Glasserman_2015} considered bankruptcy costs and their impact onto the financial system. In particular, they studied how financial organizations can in turn fail and drag other organizations to bankruptcy as a result of these costs. Simultaneously, cross-holdings were considered by Elsinger \cite{Elsinger_2009}, Elliott \emph{et al.} \cite{Elliott_2014}, Fischer \cite{Fischer_2014} and Karl and Fischer \cite{Karl_2014}. These works considered a significant aspect, namely, that cross-holdings inflate the value of the financial system and thus the net value of each organization needs to be adjusted by a factor that preserves the real value in the system \cite{Brioschi_1989}. The work by Weber and Weske integrates both aspects into a system that is able to capture fire sales as well \cite{Weber_2017}.

A crucial fact that was highlighted in these works is the inability of organizations to fulfill their obligations towards other organizations in the network that are linked via mutual liabilities, which may cause cascading defaults \cite{Elliott_2014}. The work by Birge \cite{Birge_2021} has recently proposed an inverse optimisation approach based on the decisions from national debt cross-holdings to address the propagation and extent of failures in the network. On another front, a body of literature has addressed the unrealistic assumption that payments are instantaneous and take place at the same time, e.g., \cite{Acemoglu_2015, Cabrales_2017, Chen_2021, Glasserman_2016}, proposing time-dynamic extensions of this model.

The work by Calafiore \emph{et al.} considers the problem of mitigating the financial contagion through some targeted interventions with the aim of mitigating the effects of cascading failures in the system. They consider a multi-step dynamic model of clearing payments and introduce an external control term that represents corrective cash injections made by a ruling authority \cite{Calafiore_2022}. An extension to this work considered \emph{pseudo defaults} in a dynamical setting, allowing organizations to fulfill their liabilities through dynamic clearing payments \cite{Calafiore_2023}. Similarly, a case study on the Korean financial system is proposed by Ahn and Kim where the authors study the interventions in the form of liquidity injection into the financial system under economic shocks \cite{Ahn_2019}. Finally, a recent work by Ramirez \emph{et al.} investigated the steady-state solutions of a stochastic discrete-time model where the authors study the mean and covariance error \cite{Ramirez_2022}.

Situations where companies can attain the same equity value are also of interest. This democratic view of the market is motivated by the seminal research work by Fehr and Schmidt on inequity aversion in behavioral economics~\cite{Fehr_1999}. Differently from contract-theoretic models where the underlying assumption is that agents are considered purely selfish, Fehr and Schmidt introduce a framework where psychological considerations based on empirical evidence are also integrated, e.g., fairness concerns and reciprocity~\cite{Guan_2008}. More recently, the rise of populist parties in Western democracies has led to the development of equilibrium models where voters dislike inequality \cite{Pastor_2021}. In these models, economic growth exacerbates inequality due to heterogeneity in preferences and, as a result, also in returns on capital. 

\textit{Contribution}. The contribution of this work is threefold. Firstly, we characterize the region of attraction of the equilibrium points and investigate the invariance of the system. We also consider the uncertain case where the coefficients of the matrix of cross-holding can take values in a range. Second, we consider periodic solutions and show that such solutions cannot exist for period of length $2$. However, through a counterexample, we highlight the difference between continuous-time and discrete-time monotone systems and the fact that periodic movements can be present in the latter. Lastly, we analyze the situation where the system is not in the maximal positive invariant region and investigate the optimization problem of finding the minimal cash injection that would bring the system state to that region.

The paper is organized as follows. First, we introduce the notation. In Section~\ref{sec:problem}, we present the networked model. In Section~\ref{sec:preliminary}, we provide a summary of preliminary results that are important for the rest of the paper on the existence, uniqueness and stability of the equilibrium points. In Section~\ref{sec:invariance}, we provide conditions for the invariance of each orthant and characterize the regions of attraction. In Section~\ref{sec:cycles}, we show that in general the system is not convergent to an equilibrium, but that there could exist periodic solutions of period greater than $2$. In Section~\ref{sec:invest}, we consider the problem where some organizations have failed and we wish to find the minimal investment which brings the system to the maximal positive invariant region. Finally, in Section~\ref{sec:conc}, we discuss concluding remarks and future directions.

\medskip
\noindent \textbf{Notation}. 
The symbols $\mathbb{0}_n$ and $\mathbb{1}_n$ denote the $n$-dimensional column vector with all entries equal to 0 and to 1, respectively. The identity matrix of order $n$ is denoted by $I_n$. The notation $x \ge 0$ for a generic vector $x$ or $M \ge 0$ for a generic matrix $M$ is to be intended element-wise. Given a generic vector $x \in \mathbb R^n$, let the operator $y = \phi(x)$ be such that the $i$th component $y_i = 1$ if $x_i < 0$ and $y_i = 0$ otherwise. The characteristic vector $\phi^{[k]}$ is the binary representation of integer $k$, with $k=0,1,2,\cdots, 2^n-1$. Let us introduce the diagonal matrix $J^{[k]} := {\rm diag} (\mathbb 1_n - 2\phi^{[k]})$, where the main diagonal is given by the entries of vector $\mathbb 1_n - 2\phi^{[k]}$; we denote the generic orthant $k$ by $\mathcal X^k$, namely, $\mathcal X^k := \{x \in \mathbb R^n | J^{[k]}x \ge 0 \}$.

A square real matrix $M \in \mathbb R^{n\times n}$ is said to be \emph{Metzler} if its off-diagonal entries are nonnegative, namely, $M_{ij} \ge 0$, $i \neq j$. Every Metzler matrix $M$ has a real dominant eigenvalue $\lambda_F(M)$, which is referred to as \emph{Frobenius eigenvalue}. The corresponding left and right vectors associated with $\lambda_F(M)$ are referred to as left and right \emph{Frobenius eigenvectors}, respectively \cite{Farina_2000}. 
A square real matrix is said to be \emph{Schur} stable if all its eigenvalues have absolute value less than one. Symbol $\otimes$ denotes the Kronecker product.

\section{Problem Formulation}\label{sec:problem}
In this section, we present our model \cite{Stella_2024}. This model describes a networked financial system, where the network captures the financial interdependencies between organizations. We consider a set of organizations $N = \{1, \dots, n\}$, where each organization $i \in N$ is described by an equity value $V_i \in \mathbb R$. The equity value represents the shares of the organization. 
Organizations can also hold shares of other organizations through the matrix of cross-holdings $C$, where each element of the matrix $C_{ij} \ge 0$ represents the fraction of organization $j$ owned by organization $i$ for any pair of organizations $i,j \in N$. This matrix can be seen as a network, where each row represents an organization and each element of a given row represents the amount of the shares held by this organization. Organizations can also invest in primitive assets that generate a net flow of cash over time. Let the set of primitive assets be $M = \{1, \dots, m\}$, where the market price of asset $k$ is denoted by $p_k$ and the share of the value of asset $k$ held by organization $i$ by $D_{ik} \ge 0$.

In matrix form, the following discrete-time dynamical system describes the time evolution of the equity values:
\begin{equation}\label{eq:model}
V(t+1)=CV(t)+Dp-B\phi(V(t)-\ubar V),
\end{equation}
where $t \in {\mathbb Z}^+$, $C$ is a nonnegative and nonsingular matrix for which $C_{ii}=0$ and $\mathbb{1}_n^\top  C< \mathbb{1}_n^\top$. The physical meaning of this constraint is that the equity value of each organization held by other organizations cannot exceed the total equity value of the organization itself. Matrix $D$ is a positive matrix, $p$ a nonnull nonnegative vector, $B={\rm diag}(\beta)$ a nonegative diagonal matrix with entries $\beta_i>0$, $i \in N$, $\ubar V$ is the vector of threshold values $\ubar V_i$ for all $i$ below which organization $i$ incurs a failure cost $\beta_i$ and $\phi(V-\ubar V)$ the vector of indicator functions taking value $1$ if $V_i<\ubar V_i$ and $0$ if $V_i\ge \ubar V_i$. In the above system, the first term represents the cross-holdings, the second term the primitive assets held by each organization and the last term captures the endogenous discontinuity resulting from failure.

Let $x(t) := V(t) - \ubar V$, we can rewrite system~\eqref{eq:model} as:
\begin{equation}\label{eq:modelx}
\begin{array}{c}
x(t+1) := Cx(t) + r - B \phi(x(t)), \\
r := (C - I_n) \ubar V + Dp.
\end{array}
\end{equation}
Note that the above is a monotone system since $\phi(y) \ge \phi(x)$ if $y \le x$. As a matter of fact, if $x'(0) > x''(0)$ are two initial conditions, then $x'(t) \ge x''(t)$, for any $t \ge 0$.

\section{Preliminary Results}\label{sec:preliminary}
In this section, we provide a summary of preliminary results that are used in the rest of the paper, adapted from \cite{Stella_2024}. In particular, we provide specific conditions for the positivity of system~(\ref{eq:model}) and explicitly calculate its equilibria. Finally, we consider system~(\ref{eq:modelx}) and determine necessary and sufficient conditions for the stability of the equilibria in the positive and negative orthant.

Thanks to the monotonicity of system~(\ref{eq:model}), and by recalling that $C$ is nonnegative, it is straightforward to show the following result from the condition $0\le B\phi(V(t)-\ubar V)\le \beta$.

\medskip\noindent
\begin{lemma}\label{lem1}
Consider system~(\ref{eq:model}). $V(t)\ge 0,  \forall t\ge 0$ and $V(0) \ge \mathbb 0_n$ if and only if 
\begin{equation}\label{eq.2}
Dp-\beta\ge 0.
\end{equation}  \hfill $\square$
\end{lemma}

Under condition (\ref{eq.2}), system (\ref{eq:model}) is a positive nonlinear switched system since vector $\phi(V(t)-\ubar V)$ can take a finite number of values $\phi^{[k]}$, with $k=0,1,2,\cdots, 2^n-1$. As such, system~(\ref{eq:model}) may possess at most $2^n$ equilibria in total. For instance, with $n=2$ we have:
\begin{equation}\nonumber
\begin{array}{c}
\phi^{[0]}= \mathbb{0}_n, \; \phi^{[1]}=\left[\begin{array}{c} 0 \\ 1\end{array}\right], \; \phi^{[2]}=\left[\begin{array}{c}  1 \\ 0\end{array}\right], \; \phi^{[3]}=\mathbb 1_n.
\end{array}
\end{equation}

The equilibria in orthant $k$, denoted by $\overline V^{[k]}$ and characterized by the index $k$ is given by 
\begin{equation}\label{eq.3}
\overline V^{[k]}=(I_n-C)^{-1}(Dp-B\phi^{[k]}), \quad s.t. \quad \phi(\overline V^{[k]}-\ubar V)=\phi^{[k]}.
\end{equation} 
Note that $V=0$ cannot be an equilibrium of the system since $Dp > 0$ and that, if (\ref{eq.2}) holds, $\overline V^{[k]} > 0$. In the $k$th orthant the difference $Y^{[k]}(t)=V(t) - \overline V^{[k]} = x(t) - \overline x^{[k]}$ follows the autonomous dynamics 
\begin{equation}\label{eq.4}
Y^{[k]}(t+1)=CY^{[k]}(t).
\end{equation}
Because of the nonnegativity of $C$, given $\mathbb{1}_n^\top  C< \mathbb{1}_n^\top$, therefore $C$ is Schur-stable. As a result, we can now state the following lemma whose proof is straightforward by considering the Lyapunov function $\mathcal V(x) = \mathbb 1_n^\top |x - \overline x^{[k]}| = \|x - \overline x^{[k]}\|_1$. 


\medskip\noindent
\begin{lemma}\label{lem2}
Any equilibrium $\overline x^{[k]}$ in the interior of $\mathcal X^k$ is locally asymptotically stable. \hfill $\square$
\end{lemma}
\medskip

\textit{Remark}. Note that there could be equilibrium points on the discontinuity points, but these are fragile (unstable) and are not considered in our analysis.

\medskip
We now turn our attention to the existence and uniqueness of the equilibrium points in orthants $0$ (the first orthant, where all organizations are healthy) and $2^n-1$ (the last orthant where all organizations fail), which we henceforth refer to as \emph{positive} and \emph{negative} equilibrium points, respectively. To this aim, we consider system~\eqref{eq:modelx} and recall the following result.

\medskip
\begin{lemma}\label{lem3}
Consider system~\eqref{eq:modelx}. In each open orthant $\mathcal X^k$, there exists at most one equilibrium. Furthermore, the following points hold true:
\begin{enumerate}
\item There exists an equilibrium point $\bar x \ge 0$ if and only if $(I_n - C)^{-1}r \ge 0$.
\item If  $(I_n - C)^{-1}(r - \beta) \ge 0$, then there exists an equilibrium point $\bar x \ge 0$ and it is the unique equilibrium.
\item There exists an equilibrium point $\bar x < 0$ if and only if $(I_n - C)^{-1}(r - \beta) < 0$.
\item If  $(I_n - C)^{-1}r < 0$, then there exists an equilibrium point $\bar x < 0$ and it is the unique equilibrium.
\end{enumerate} \hfill $\square$
\end{lemma}
\textit{Proof}. First, let us prove the first statement, namely, if an equilibrium exists in orthant $k$, it is unique. Let 
$$\bar x^{[k]} = (I_n - C)^{-1}(r - B \phi^{[k]}) \in \mathcal X^k$$
be the generic equilibrium point in othant $k$. By contradiction, let us assume that a second equilibrium point exists in the same orthant. It is straightforward to see that the calculation with a given $\phi^{[k]}$ would produce the same equilibrium point. \\
Let us now prove the rest point by point.
\begin{enumerate}
\item Let $(I_n - C)^{-1}r \ge 0$, then $\bar x = (I_n - C)^{-1}r \ge 0 \in \mathcal X^0$. Vice versa, assume that there exists a generic equilibrium $\bar x \ge 0$, then $\bar x \in \mathcal X^0$. Therefore, $\phi(\bar x) = 0$ and $(I_n - C)^{-1}r \ge 0$.
\item Let $(I_n - C)^{-1}(r - \beta) \ge 0$, then $(I_n - C)^{-1}r \ge (I_n - C)^{-1} \beta \ge 0$. It follows from the first point that there exists an equilibrium $\bar x \ge 0$. Moreover, assume there exists an equilibrium $\bar x^{[k]}$ in orthant $\mathcal X^k$, i.e., $\bar x^{[k]} = (I_n -C)^{-1}r - B \phi(x^{[k]}) \ge (I_n -C)^{-1}(r - \beta) \ge 0$. Then, the unique equilibrium is in orthant $\mathcal X^0$. 
\item Let $(I_n - C)^{-1}(r - \beta) < 0$, then $\bar x = (I_n - C)^{-1}(r - \beta) < 0 \in \mathcal X^{2^n-1}$. Vice versa, assume that there exists a generic equilibrium $\bar x < 0$, then $\bar x  \in \mathcal X^{2^n-1}$. Therefore, $\bar x = (I_n - C)^{-1}(r - \beta) < 0$.
\item Let $(I_n - C)^{-1}r \le (I_n - C)^{-1}(r - \beta) < 0$, then from point 3, there exists an equilibrium $\bar x^{[k]} < 0$. Moreover, assume there exists an equilibrium $\bar x^{[k]}$ in orthant $\mathcal X^k$, i.e., $\bar x^{[k]} = (I_n -C)^{-1}r - B \phi(x^{[k]}) \le (I_n -C)^{-1}r < 0$. Then, the unique equilibrium is in orthant $\mathcal X^{2^n-1}$. 
\end{enumerate}
This concludes our proof. \hfill $\blacksquare$ \\

In the following section, we build on these preliminary results to investigate invariance and the existence of a region of attraction in a generic orthant $k$.

\section{Region of Attraction and Invariance}\label{sec:invariance}
In the previous section, we have characterized the equilibria of systems~\eqref{eq:model}-\eqref{eq:modelx} and provided conditions for the existence and uniqueness of such equilibria. Moreover, we have shown that all possible equilibria are locally asymptotically stable, so that we are now in a position to investigate the region of attraction of equilibria in a generic orthant $k$ \cite{Blanchini_1999}. To this end, let us introduce the diagonal matrix $J^{[k]}$, with diagonal entries $J^{[k]}_{ii} = 1$ if $i$ is such that $V_i \ge \ubar V_i$, and $J^{[k]}_{ii} = -1$ if $i$ is such that $V_i < \ubar V_i$.

\medskip
\begin{theorem}\label{th1}
Consider system~\eqref{eq:model}. The region of attraction in orthant $k$ for equilibrium $\overline V^{[k]}$ is given by all the initial values $V(0) = V^{[k]}(0)$ satisfying
\begin{equation}\label{eq:region}
J^{[k]}(C^t(V(0) - \overline V^{[k]}) + \overline V^{[k]} - \ubar V) \ge 0, \quad t=0,1,2,\dots
\end{equation}  \hfill $\square$
\end{theorem}

\medskip\noindent

\textit{Proof}. We have to characterize the invariant set of $V(0)$ in orthant $k$ such that $V(t)$ lies in the same orthant and so eventually converges to $\overline V^{[k]}$ (thanks to Lemma \ref{lem2}). As such, consider the dynamics in orthant $k$, i.e., from (\ref{eq.4})
$$
V(t) - \ubar V = \overline V^{[k]} - \ubar V + C^t(V^{[k]}(0) - \overline V^{[k]}).
$$
$V(t)$ in orthant $k$ is equivalent to 
$$
J^{[k]}(V(t) - \ubar V) \ge 0, \quad t=0,1,2,\dots
$$
This concludes our proof. \hfill $\blacksquare$

\medskip\noindent
%

\medskip\noindent
\medskip\noindent
We are now interested in characterizing the first orthant $\mathcal X^0$, namely, $k = 0$, where all organizations are healthy, and the last orthant $\mathcal X^{2^n-1}$, i.e., $k = 2^n-1$, where all organizations fail. To this end, we introduce the necessary and sufficient conditions for the invariance of orthants $\mathcal X^0$ and $\mathcal X^{2^n-1}$. The next result provides a sufficient and necessary condition for orthant $\mathcal X^0$ to be invariant.

\begin{theorem}\label{th2}
Orthant $\mathcal X^0$ is invariant if and only if 
\begin{equation}\label{eq:ort0}
(C - I_n) \ubar V + Dp \ge 0.
\end{equation} 
\end{theorem}

\textit{Proof}. Consider the difference $x(t) = V(t) - \ubar V$ that obeys to the following difference equation
$$
x(t+1)=C x(t) + (C - I_n) \ubar V + Dp - B \phi(x(t)).
$$
From the condition in the statement of the theorem, we have that 
$$
x(t+1) \ge C x(t) - B \phi(x(t)).
$$
Therefore, $x(0) \ge 0$ yields $\phi(x(0))=0$ (orthant $\mathcal X^0$), such that, thanks to the nonnegative property of $C$, also $x(1) \ge 0$. The same reasoning can be used for $x(2)$ and so on. This concludes the proof. \hfill $\blacksquare$

The next result provides a sufficient and necessary condition for orthant $\mathcal X^{2^n-1}$ to be invariant.

\medskip\noindent
\begin{theorem} \label{th3}
Orthant $\mathcal X^{2^n-1}$ is invariant if and only if the following condition holds true:
\begin{equation}\label{eq:ortneg}
(C - I_n) \ubar V + Dp < \beta.
\end{equation} 
\end{theorem}
\textit{Proof}. Consider again the difference $x(t) = V(t) - \ubar V$ that obeys to following equation
$$
-(x(t+1)) = C(-x(t)) - (C - I_n) \ubar V - Dp + B \phi(x(t)).
$$
Thanks to the condition in the statement of the theorem, we have that 
$$ 
-x(t+1) \ge C(-x(t)) + B \phi(x(t)) - \beta.
$$
Therefore, $-x(0) \ge 0$ yields $\phi(x(0))= {\mathbb 1}_n$ (in orthant $\mathcal X^{2^n-1}$), such that, thanks to the nonnegative property of $C$, it follows that also $-x(1)\ge 0$. This applies to the next orthant as well in a similar manner to the previous theorem. This concludes the proof. \hfill $\blacksquare$


We can now conclude that, for $k \ne 0, k \ne 2^n-1$, the generic orthant $k$ cannot be invariant, whereas this is possible for $k=0$ and $k=2^n-1$. This fact is proved in the following theorem.

\medskip\noindent
\begin{theorem}\label{th4}
Let $C_{ij}>0$, for any $i \ne j$. Then, every orthant $k$ with $k \ne 0$ and $k \ne 2^n-1$ is not invariant.
\end{theorem}

\textit{Proof}. Assume by contradiction that a generic orthant $k$ is invariant. This means that inequality (\ref{eq:region}) holds true for any $V(0)$ in a generic orthant $k$ and any $t \ge 0$. Let $t=1$, $y=J^{[k]}(V(0) - \ubar V) \ge 0$, $d = J^{[k]}(I_n - C)(V^{[k]} - \ubar V)$, then inequality (\ref{eq:region}) reads 
$$
J^{[k]} C J^{[k]}y + d \ge 0.
$$
Since $k \ne 0$ and $k \ne 2^{n-1}$, there exist two integers, namely, $i$ and $j$, with $J^{[k]}_{ii} = -1$ and $y_j = V^{[k]}_j - \ubar V_j > 0$. Then, row $h$ of the above inequality reads
$$
-\sum_{h \ne i,j} C_{ih} y_h + d_h - C_{ij} y_j > 0,
$$ 
for any values of $y_j > 0$, i.e., for any values of $V^{[k]}_j(0) > \ubar V_j$.
Since $C_{ij} > 0$, we have arrived to a contradiction. This concludes our proof. \hfill $\blacksquare$

\medskip
In the next result, we show that the invariant sets in the positive and negative orthants are finitely determined, which cannot be stated for the other orthans as discussed in the remark that follows.

\medskip\noindent
\begin{theorem}\label{th5}
In orthants $\mathcal X^0$ and $\mathcal X^{2^n-1}$, the invariant sets are finitely determined, i.e., they can be computed through Theorem \ref{th1}, namely, inequalities (\ref{eq:region}), for $t=1,2,\dots, \tau$, where $\tau$ is the minimum integer for which 
\begin{eqnarray}\label{eq:invsets}
(C^\tau (\ubar V - \overline V^{[k]}) + \overline V^{[k]} - \ubar V) \ge 0, \quad k&=&0, \\
\label{eq:invsets2}
(C^\tau (\ubar V - \overline V^{[k]}) + \overline V^{[k]} - \ubar V) \le 0, \quad k&=&2^{n}-1.
\end{eqnarray} 
\end{theorem}

\textit{Proof}. Let $\overline x = \overline V - \ubar V$ and consider orthant $k=0$, i.e., $\overline x \ge 0$ (for the other orthant the proof is similar). Define the set 
$$
{\cal P}_\tau = \{ x \quad s.t. \quad C^h x + \overline x \ge 0, \quad t=0,1,2,\dots, \tau \}.
$$
Clearly, it results that ${\cal P}_{\tau+1} \subset {\cal P}_\tau$. Assume that for $h=\tau$, inequality (\ref{eq:invsets}) holds, and take $x \in {\cal P}_\tau$. Then $C^{\tau+1} x + \overline x = C^{\tau}C x + \overline x \ge C^\tau (- \overline x) + \overline x \ge 0$, i.e., $x \in {\cal P}_{\tau+1}$. Therefore, ${\cal P}_\tau  \subset {\cal P}_{\tau+1}$. In conclusion ${\cal P}_\tau = {\cal P}_{\tau+1}$. This concludes our proof. \hfill $\blacksquare$

\medskip\noindent
\textit{Remark}. Note that for $\tau=1$ in Theorem \ref{th5} the necessary and sufficient conditions for invariance of the orthants $\mathcal X^0$ and $\mathcal X^{2^n-1}$ are recovered, see Theorems \ref{th2} and \ref{th3}. The fact that the invariant sets for orthants $\mathcal X^0$ and $\mathcal X^{2^n-1}$ are finitely determined (Theorem \ref{th5}) cannot be proved for the other orthants $k=1,2,\cdots, 2^n-2$. This is due to $J^{[k]}$ being not sign definite, so that $J^{[k]} C^h J^{[k]}$ is no longer a positive matrix.

\medskip\noindent
\begin{example}\label{ex.1}
Consider the following toy example for system (\ref{eq:model}) with the matrix of cross-holdings $C$, matrix $D$, vector $p$, vector $\beta$ and threshold $\ubar V$ set as: 
\begin{eqnarray}\nonumber
& C=\left[\begin{array}{cc} 0 & 0.5\\ 0.5 & 0\end{array}\right], \quad D=\left[\begin{array}{cc} 0.5 & 0.25\\ 0.25 & 0.5\end{array}\right], \\ 
\nonumber & p=4\mathbb 1_2, \quad \beta = \mathbb 1_2, \quad \ubar V = 5 \mathbb 1_2.
\end{eqnarray}
From the above parameters, we know that the first and third quadrants are invariant in accordance with Theorems \ref{th2} and \ref{th3}. In quadrants 2 and 4, the invariant sets are $4 < V_1 < 5$, $5 < V_2 < 6$ and $5 < V_1 < 6$, $4 < V_2 < 5$, respectively, in accordance with Theorem \ref{th1}. Figure~\ref{fig:ex1pp} illustrates the phase portrait of the system in the $V_1$-$V_2$ plane, where each quadrant has an equilibrium point, indicated by a red circle, and the invariant sets in quadrants 2 and 4 are indicated by thick black rectangles.

\begin{figure}[t]
        \centering
        \includegraphics[width=.45\textwidth]{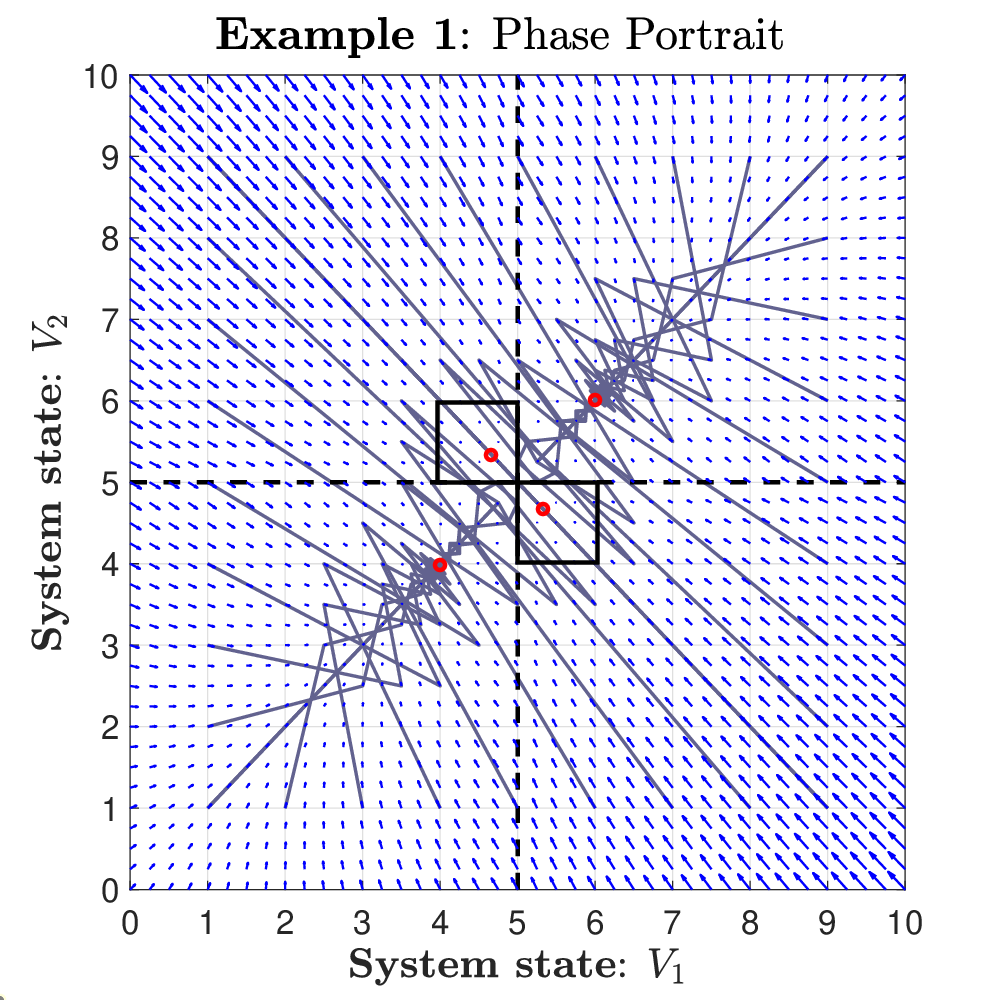}
        \caption{Phase portrait in the $V_1$-$V_2$ plane. The equilibria are indicated by red circles and the thick black rectangles denote the invariant sets in quadrants 2 and 4.}
        \label{fig:ex1pp}
\end{figure}
\end{example}


\subsection{The Uncertain Case}
We now consider the case where the matrix of cross-holdings is time-variant, namely, $C(t)$, and derive a result on the largest invariant set of the positive orthant, i.e., $\mathcal X^0$. We wish to consider the positive orthant, i.e., the space where all organizations are healthy $x \geq 0$, so we assume that the model is of the form:
 \begin{equation}
\label{diff_inclu}
x(t+1) = C(t) x(t) + r,~~~0 < c_{ij}^- \leq c_{ij}(t) \leq c_{ij}^+,
  \end{equation}
where $c_{ij}^-$ and $c_{ij}^+$ are given bounds.

Define as $C^+$ and $C^-$ as the matrices for which $c_{ij} = c_{ij}^-$ and
$c_{ij} = c_{ij}^+$, respectively, and define 
$$
\bar x^- = (\mathbf I - C^-)^{-1} r,
$$
$$
\bar x^+ = (\mathbf I - C^+)^{-1} r.
$$
Then, we can state the following result.

\begin{theorem}\label{th7}
Consider system \eqref{diff_inclu}. Then
\begin{itemize}
\item 
The largest robust invariant set ${\cal R}$ inside the healthy set for \eqref{diff_inclu}
is the largest invariant set ${\cal P}^-$ for 
$$
x(t+1) = C^- x(t) + r.
$$
\item
For any initial condition in ${\cal R}$, the system converges to the hypercube
$\bar x^- \leq x \leq \bar x^+$:
$$
\limsup_{t \rightarrow \infty} x(t) \leq \bar x^+
$$
and
$$
\liminf_{t \rightarrow \infty} x(t) \geq \bar x^-.
$$
\item
The largest robust invariant set ${\cal P}^+$ for 
$$
x(t+1) = C^+ x(t) + r
$$
is the last-hope set. For any initial condition $x(0) \geq 0$
$x(0) \not\in {\cal P}^+$ the state vector will become non-positive
at some instant $t$ (some organization will fail) no matter how 
$c_{ij}(t)$ evolves.
\end{itemize}
\end{theorem}
{\it Proof}. The largest robust invariant set ${\cal R}$ is a subset of ${\cal P}^-$, the largest 
invariant set for $x(t+1) = C^- x(t) + r$, because $c_{ij}(t)=c_{ij}^-$
is a possible realization. Hence ${\cal R} \subseteq {\cal P}^-$. 

To show that the two sets are equal,
take any $x(0) \in {\cal P}^-$. We prove that $x(0) \in {\cal R}$, so ${\cal P}^- \subseteq {\cal R}$.
Take any sequence $C(t)$. Let $x(t)$ be the solution and $x^-(t)$ be the solution with $c_{ij}(t)=c_{ij}^-$. We have
$$
x(1) = C(1) x(0) + r \geq x(1) = C^- x(0) + r = x^-(1).
$$
Now, assume  $x(t) \geq x^-(t)$. Then, recursively
\begin{align*}
    x(t+1)  = C(t) x(t) + r & \geq  C^- x(t) + r \\
    & \geq x C^- x^-(t) + r = x^-(t+1).
\end{align*}
Hence, $x(t) \geq x^-(t)$, for all $t$. On the other hand $x^-(t) \geq 0$, by construction.

\medskip\noindent
The second part of the proof is as follows. We consider the solutions $x^-(t)$ and $x^+(t)$ of the extremal systems, for the same initial condition and prove that
$$
x^-(t) \leq x(t) \leq x^+(t),
$$
which follows directly from the considerations made previously.

\medskip\noindent
The third part of the proof is also immediate. If $x(0) \geq 0$, $x(0) \not\in {\cal L}$, then $x^+(t)$ becomes negative at some point.
Since $x(t) \leq x^+(t)$, this concludes our proof. \qed

\section{Periodic Solutions}\label{sec:cycles}
In this section, we look for possible periodic solutions of system (\ref{eq:model}). The importance of the absence of periodic solutions lies in the fact that for any initial condition the system converges to an equilibrium point. Indeed, if the system does not admit limit cycles, then the system is {\it convergent}, according to the definition given in \cite{Blanchini_2019}. Note that both equilibria and the possible periodic movements are asymptotically stable. We can only prove that periodic movements of period $h=2$ cannot exist. Through a counterexample, we show that the system is not convergent in general, i.e., there could exist an initial state generating a stable periodic movement of period $h>2$.

\noindent
The search of periodic solutions calls for the so-called lifted system. Indeed, consider again the dynamics of the difference $x(t) = V(t) - \ubar V$ in equation~(\ref{eq:modelx}). Since we want to look for a periodic movement of $x(t)$ of period $h > 1$, for a certain $t = 0,1,2, \dots$, let 
$$
Z = \left[\begin{array}{c} X(t) \\ X(t+1) \\ \vdots \\ X(t+h-2) \\ X(t+h-1)\end{array}\right],
$$ 
with the constraint $X(t+h)=X(t)$. From (\ref{eq:invsets}), we can write
\begin{equation}\label{eq:Z}
Z = \tilde C Z+{\mathbb 1}_n\otimes (Dp + (C - \mathbf I) \ubar V) - \tilde B \phi(Z),
\end{equation}
where 
\begin{scriptsize}
$$
\tilde C = \left[\begin{array}{ccccc} 0 & 0 &  \cdots  & 0 & C\\ C & 0 & \cdots & 0 & 0\\ 0 & C & \cdots & \cdots & 0\\ \vdots & \vdots & \ddots & \vdots & \vdots\\ 0 & 0 & \cdots & C & 0  \end{array}\right], \quad
\tilde B = \left[\begin{array}{ccccc} 0 & 0 &  \cdots  & 0 & B\\ B & 0 & \cdots & 0 & 0\\ 0 & B & \cdots & \cdots & 0\\ \vdots & \vdots & \ddots & \vdots & \vdots\\ 0 & 0 & \cdots & B & 0  \end{array}\right],
$$
\end{scriptsize}
and $\phi(Z)$ is a column with column entries $\phi(X(t+i))$, $i=0, 1, \dots, h-1$.  Note that equations (\ref{eq:Z}) have solutions $\bar Z={\mathbb 1}_n \otimes (V^{[k]} - \ubar V)$, where $V^{[k]}$ is the (possible) equilibrium of the system in orthant $k$. Looking for possible cycles of length $h$ can be interpreted as finding non trivial solutions (i.e. $Z \ne \bar Z={\mathbb 1}_n\otimes (V^{[k]} - \ubar V))$ of (\ref{eq:Z}). Let $W = Z - \ubar Z$, then we have 
$$
Z - \bar Z = P(\phi (\bar Z) - \phi (Z)), \quad P=(\mathbf I - \tilde C)^{-1} \tilde B.
$$

\medskip\noindent
In the following result, we prove the nonexistence of cycles of length $2$. 

\medskip
\begin{theorem}\label{th6}
System~\eqref{eq:modelx} does not possess periodic movements with period $2$.
\end{theorem}

\textit{Proof}. Let $h=2$ in system (\ref{eq:Z}). This implies that $x(2) = x(0)$ and, therefore:
\begin{eqnarray*}
x(0) = C x(1) + r - B\phi(x(1)), \\
x(1) = C x(0) + r - B\phi(x(0)).
\end{eqnarray*}
By subtracting the second equation to the first, we get $x(0) - x(1) = - C(x(0) - x(1)) + B(\phi(x(0)) - \phi(x(1)))$, which can be rewritten as:
$$
\xi = - C \xi + B\zeta,
$$
where $\xi := x(0) - x(1)$, $\zeta := \phi(x(0)) - \phi(x(1))$. 
Define the diagonal matrix $S$, whose diagonal entry $S_{ii}$ is equal to 1 if $\xi_i \ge 0$ and $S_{ii}$ is equal to 0, otherwise. Notice that $S \zeta \le 0$. Therefore,
$$
(2S - I_n) \xi = |\xi| \le -(2S - I_n)C \xi \le C |\xi|,
$$
which is impossible, since $C$ is a positive matrix and $\mathbb 1_n^\top C < \mathbb 1_n^\top$. Indeed, from the above inequality we have that $\mathbb 1_n^\top |\xi| \le \mathbb 1_n^\top C |\xi| < \mathbb 1_n^\top |\xi|$, which is a contradiction. This concludes our proof. \hfill $\blacksquare$

\medskip
\noindent One is tempted to conjecture that, due to the special structure of the system, no limit cycles exist for periods of any length. This is not true in general, as shown in the following counterexample. 

\medskip
\begin{example}\label{ex.controesempio}
Consider system (\ref{eq:modelx}) with 
\begin{eqnarray*}
C&=&\left[ \begin{array}{cccc} 0 & 0 & 0 & 0.8\\ 0.8 & 0 & 0 & 0\\ 0 & 0.8 & 0 & 0\\ 0 & 0 & 0.8 & 0\end{array}\right], \\
\ubar V&=& 7.5\mathbb 1_4, \quad D=0.5I_4, \quad p=5\mathbb 1_4, \quad B=2I_4.
\end{eqnarray*}
Letting $x=V-\underbar x$ the system becomes
$$
x(t+1)=Cx+\mathbb 1_4-2\phi(x).
$$
Note that the condition of Lemma \ref{lem1} holds true, and so are the conditions of Lemma \ref{lem3}-1) and Lemma \ref{lem3}-3). Indeed, the system admits one equilibrium in the positive orthant, i.e., $\bar x^{[0]}=5\mathbb 1_4$, one in the negative orthant, i.e. $\bar x^{[15]}=-5\mathbb 1_4$ and other 6 equilibria, i.e., 
$$
\pm \left[ \begin{array}{c} \alpha\\ \gamma\\ -\alpha\\ -\gamma\end{array}\right], \quad \pm\delta\left[ \begin{array}{c} 1\\ -1\\1\\-1\end{array}\right], \quad  \pm\left[ \begin{array}{c} \gamma\\ -\alpha\\ -\gamma\\ \alpha\end{array}\right], 
$$
with $\alpha=0.1220$, $\gamma = 1.0976$, $\delta=0.5556$. 

\medskip\noindent
Starting from $x(0)=[0.6754\, \; -1.3678\, \; -0.6754\, \; 1.3678]^\top$, the state of the system evolves according to a periodic trajectory of period $h=8$. Precisely, the states $x(0)$ to $x(7)$ are grouped in rows of the following matrix:
$$
X_P =\left[ \begin{array}{cccccccc}
    0.6754  & -1.3678 & -0.6754 & 1.3678 \\
    2.0942 &   -0.4597 & -2.0942 & 0.4597\\
    1.3678 &   0.6754 &  -1.3678 &  -0.6754 \\
    0.4597 & 2.0942 & -0.4597 & -2.0942\\
    -0.6754  & 1.3678 & 0.6754 & -1.3678\\
    -2.0942 &   0.4597 & 2.0942 & -0.4597\\
    -1.3678  & -0.6754 &   1.3678 &  0.6754 \\
    -0.4597 & -2.0942 & 0.4597 & 2.0942\\
    \end{array}\right].
$$

Figure~\ref{fig:controesempio} depicts the periodic trajectory of state $x(t)$ with the above parameters.


\begin{figure}[t]
    \centering
    \includegraphics[width=0.45\textwidth]{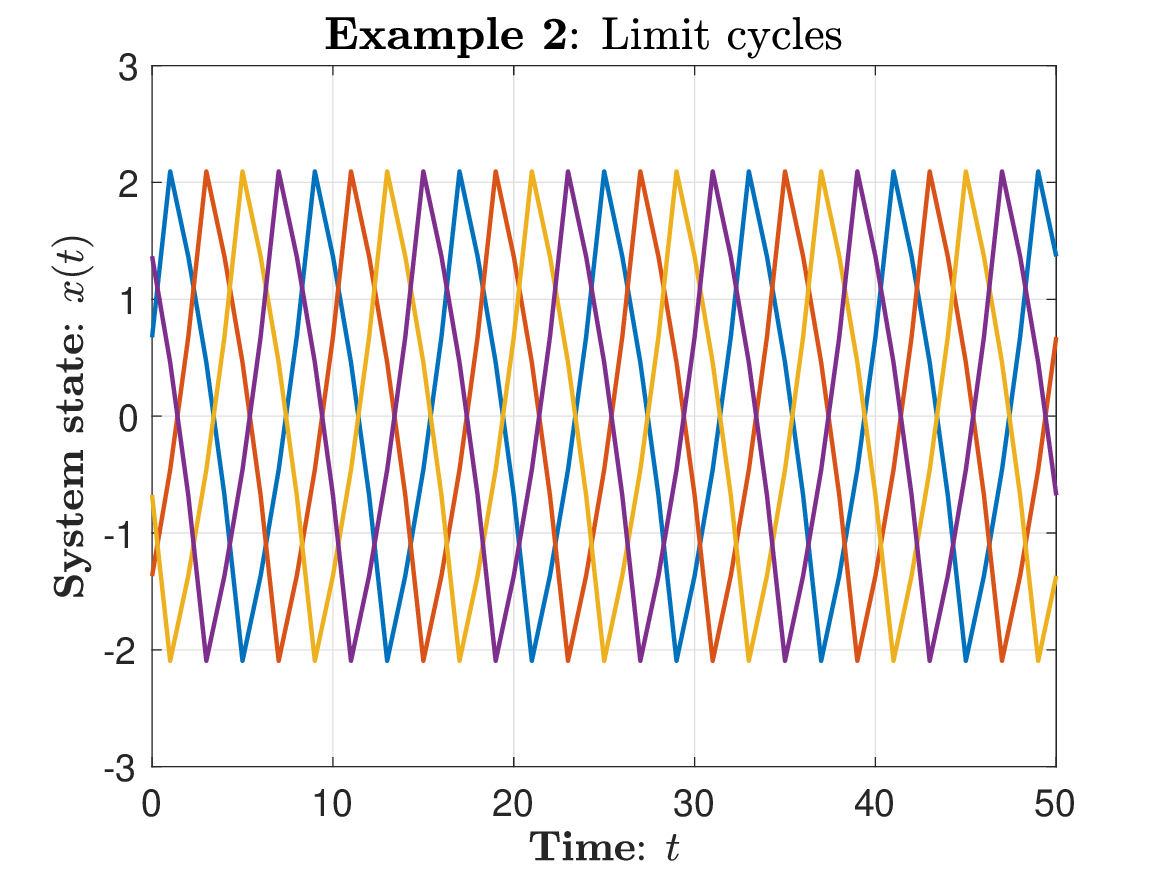}
    \caption{The trajectory of system $x(t)$, which is a periodic movement of period $8$. Each color represents a different organization.}
    \label{fig:controesempio}
\end{figure}
\end{example}

\medskip
\textit{Remark}. The above is a counterexample in the discrete-time case, whose behavior differ from the one of monotone continuous-time systems. Indeed, for monotone continuous-time systems if a periodic trajectory exists, it is unstable~\cite{Hirsh_1988, Polacik_1992, Hirsh_2006}. On the other hand, for monotone discrete-time systems, there can be periodic movements that are stable. Finally, the system can be regularized by eliminating the discontinuity, i.e., by replacing $sign(x)$ with $sat(kx)$ or $tanh(kx)$, with large $k$ while preserving this difference with continuous-time systems.

\medskip
In the following, we show that trajectories that do not fall in the discontinuities can only either converge to an equilibrium point or to a periodic solution. Before we discuss our next result, we provide the following definition.

\medskip
\begin{definition}
A trajectory is \emph{non-critical} if $|x_k(t)| \geq \rho$, for some (arbitrarily small) $\rho>0$. A (constant) vector $x$ is \emph{non-critical} if $|x_k| \geq \rho$, for all $k$.
\end{definition}
\begin{definition}
Two vectors $x_k(t)$ and $x'_k(t)$ are \emph{separated} if $x_k(t)$ and $x'_k(t)$ have at least a different sign for some $t$ and~$k$. Two vectors are \emph{separated} if they have components with different signs.
\end{definition}

Note that if two trajectories are non separated, their difference $z(t)=x_k(t)-x'_k(t)$ evolves as $z(t+1)=C z(t)$. Therefore $\|z(t)\|$ is decreasing for some norm $\|\cdot \|$.

\medskip
\begin{theorem}
Consider system~\eqref{eq:modelx}. Any non-critical trajectory $x(t)$ converges to either an equilibrium or a periodic trajectory. \hfill $\square$
\end{theorem}

\textit{Proof}. Let $x(t)$ be a non-critical trajectory for some $\rho$. Since  $x(t)$  is bounded it has an accumulation point $\bar x$. This accumulation point $\bar x$ must be necessarily non-critical.
 
Let $\bar x(t)$ the trajectory that originates at $\bar x$ at some $t_0$, $\bar x(t_0)= \bar x$.
Consider the neighbourhood 
$$
\mathcal{N}=\{x:~\|x-\bar x\|\leq \rho \}.
$$ 
Now, since $\bar x$ is an accumulation point, there exists $t_0$ such that $x(t_0) \in \mathcal{N}$, namely, $\|x(t_0) - \bar x\|\leq \rho$.

Consider the two trajectories $x(t)$ and $\bar x(t)$ for $t \geq t_0$. These two trajectories are non-separated. Indeed, at time $t_0$ their components have the same sign because if $x(t_0)$ and $\bar x(t_0)$ have different sign in some components, say $k$, then we would have
$$
|x_k(t_0)-\bar x_k(t_0)| = |x_k(t_0)-\bar x|\geq 2\rho,
$$
which is in contradiction with $x(t_0) \in \mathcal{N}$. Let $z(t) := x(t)-\bar x(t)$. Then,
$$
z(t_0+1) = C z(t_0).
$$
But this implies $\| z(t_0+1) \|=\| x(t_0+1)-\bar x(t_0+1)\| \leq \rho$. By using the same reasoning as above, this means that $x(t_0+1)$ and $\bar x(t_0+1)$ are non-separated. Recursively, one can see that $\| x(t_0+2)-\bar x(t_0+2)\| \leq \rho$, and that $x(t_0+2)$ and $\bar x(t_0+2)$ are non-separated. Then, the following
\begin{equation}\label{zeta}
z(t+1) = C z(t)
\end{equation}
holds for $t \geq t_0$ and the two trajectories $x(t)$  and $\bar x(t)$ are non-separated.
 
On the other hand, equation~\eqref{zeta} implies that $z(t) \rightarrow 0$ and so $x(t) \rightarrow \bar x(t)$. Therefore, we need to prove that $\bar x(t)$ is periodic. Given $\epsilon$ arbitrarily small, there exists $t_1$ large enough such that the following two conditions hold: first,
$$
\|x(t_1)-\bar x\| \leq \epsilon/2,
$$
because $\bar x$ is an accumulation point and, second,
$$
\|x(t_1)-\bar x(t)\| \leq \epsilon/2,
$$
because $x(t)$ converges to $\bar x(t)$. Putting these conditions together we have that
$$
\| \bar x(t_1) -\bar x \| \leq \epsilon,
$$
for $t_1$ large enough.  
 
Take $T=t_1-t_0$. Since the trajectories are non-separated, we can consider the lifted linear system
$$
z(t+T) = C^T z(t),
$$
and we since $\|z(t)\|$ is decreasing we have that, given $\bar x(t_0) \in \mathcal{N}$, it follows that $\bar x(t_0+T) \in \mathcal{N}$ as $\epsilon$ is arbitrarily small.
 
Consider the periodic system with period of length $T$ which generates both $\bar x(t)$ and $x(t)$
$$
x(t+1) = C x(t) + \psi(t),~~~\bar x(t+1) = C \bar x(t) + \psi(t),
$$ 
where the periodic signal $\psi(t) = \Psi(x(t))$ is common to both these non-separated trajectories. Lifting the systems we have for both trajectories yields
$$
\bar x(t+T) = C^T \bar x(t) + C^{T-1} \psi(t+1) + \cdots + C^{1} \psi(T-1) +\psi(t).
$$
Since $\epsilon$ is arbitrary small, this $T$--lifted system has the fixed point $\bar x$.
On the other hand, by construction,  $\bar x$ is also the initial state, so $\bar x(t)$ is periodic with period of length $T$.

Finally, note that the constructed $T$ is not the smallest period, but since it is an integer, there will be a ``smallest" one, possibly $T=1$. This concludes our proof. \hfill $\blacksquare$

\section{Minimal Cash Injection for a Healthy System}\label{sec:invest}
In this section, we study the case where the system state in not in the positive orthant, namely, not all the organizations are healthy. We are interested in finding the minimal cash injection that would bring the system to the maximal positive invariant region.

We consider the problem of finding the minimal investment such that we retain the maximal invariant region of the positive orthant. In detail, we define this problem as
\begin{align}
    \label{eq:min1} & \min_v \mathbb 1_n^\top v, \\
    \label{eq:min2} & {\rm s.t.} \quad x + v \in \mathcal M^+,
\end{align}
where $v$ is the minimal cash injection into the system that brings the system in $\mathcal M^+$, namely, the maximal invariant region of the positive orthant. This investment can be seen as a Kronecker delta function to the state. For instance, a sample initial state $x(0)$ and its corresponding minimal cash injection $v$ are:
$$x(0) = \left[ \begin{array}{c} -0.2943 \\
   -1.0177 \\
   -0.0024 \\
    0.4985 \\
   -0.0982 \\
    0.0954 \\
   -0.0425 \\
    0.4426 \\
   -0.7542 \\
   -1.3096 \end{array} \right], \quad v = \left[ \begin{array}{c} 0.2943 \\
    1.0177 \\
    0.0024 \\
   -0.4985 \\
    0.0982 \\
   -0.0954 \\
    0.0425 \\
    5.5322 \\
    0.7542 \\
    5.1135 \end{array} \right].$$ 

Now, we consider the following optimization problem, where we use the output of the minimization in \eqref{eq:min1}-\eqref{eq:min2} in the dynamical setting where we want to bring the system to $\mathcal M^+$. The optimization problem at each step can be formulated as:
\begin{align}
    \label{eq:opt1} &  \min_D \| Dp - v\|_2 + \| \mathbb 1_n^\top D \|_2  \\
    \label{eq:opt2} & {\rm s.t.} \quad \mathbb 1_n^\top D \le \mathbb 1_n^\top, \\
    \label{eq:opt3} & D \ge 0, \\
    \label{eq:opt4} & (I_n - C)^{-1}Dp - \ubar V > 0.
\end{align}
where the constraints are such that the columns of matrix $D$ sum to at most $1$, that the entries of matrix $D$ are nonnegative, and that the condition of Lemma \ref{lem3}-1) holds true. The pseudocode of the optimization problem is shown in Algorithm \ref{alg:opt}.

\renewcommand{\algorithmicrequire}{\textbf{Input:}}
\renewcommand{\algorithmicensure}{\textbf{Output:}}
\begin{algorithm}[h!]
\caption{Minimal Cash Injection for $x(t) \in \mathcal M^+$}
\label{alg:opt}
\begin{algorithmic}[1]
\Require $x(0)$, $v$, $C$, $p$, $\ubar V$, $B$
\Ensure $D$ such that $x(t) \in \mathcal M^+$
\State Initialize starting state: $x_0=x(0)$.
\State Compute $\mathcal M^+$ from \eqref{eq:invsets}.
\State Find $v$ in \eqref{eq:min1}-\eqref{eq:min2}.
\While{$x(t) \notin \mathcal M^+$}
\State Find $D$ in \eqref{eq:opt1}-\eqref{eq:opt4}.
\State Update state: $x(t)$ from \eqref{eq:modelx}.
\State Update investment: $v \leftarrow v - x(t)$.
\EndWhile
\end{algorithmic}
\end{algorithm}

\begin{example}
Consider system (\ref{eq:modelx}) with $n=10$ and
\begin{eqnarray*}
C&=& \frac{1}{n+2} \Big ( \mathbb 1_n \mathbb 1_n^\top - I_n \Big ), \\
\ubar V&=& 0.5\mathbb 1_n, \quad p=\mathbb 1_n, \quad B=0.4I_n.
\end{eqnarray*}

We set $x(0)$ to an initial random condition where some organizations are healthy and some have failed. We then solve the minimization problem \eqref{eq:min1}-\eqref{eq:min2} to find $v$. Then, by using Algorithm \ref{alg:opt} we obtain a minimal $D$ over time ensuring that the system $x(t)$ reaches the maximal positive invariant region $\mathcal M^+$. Figure~\ref{fig:cash} depicts the time evolution of the system under the considered optimization problem.

\begin{figure}[t]
    \centering
    \includegraphics[width=.45\textwidth]{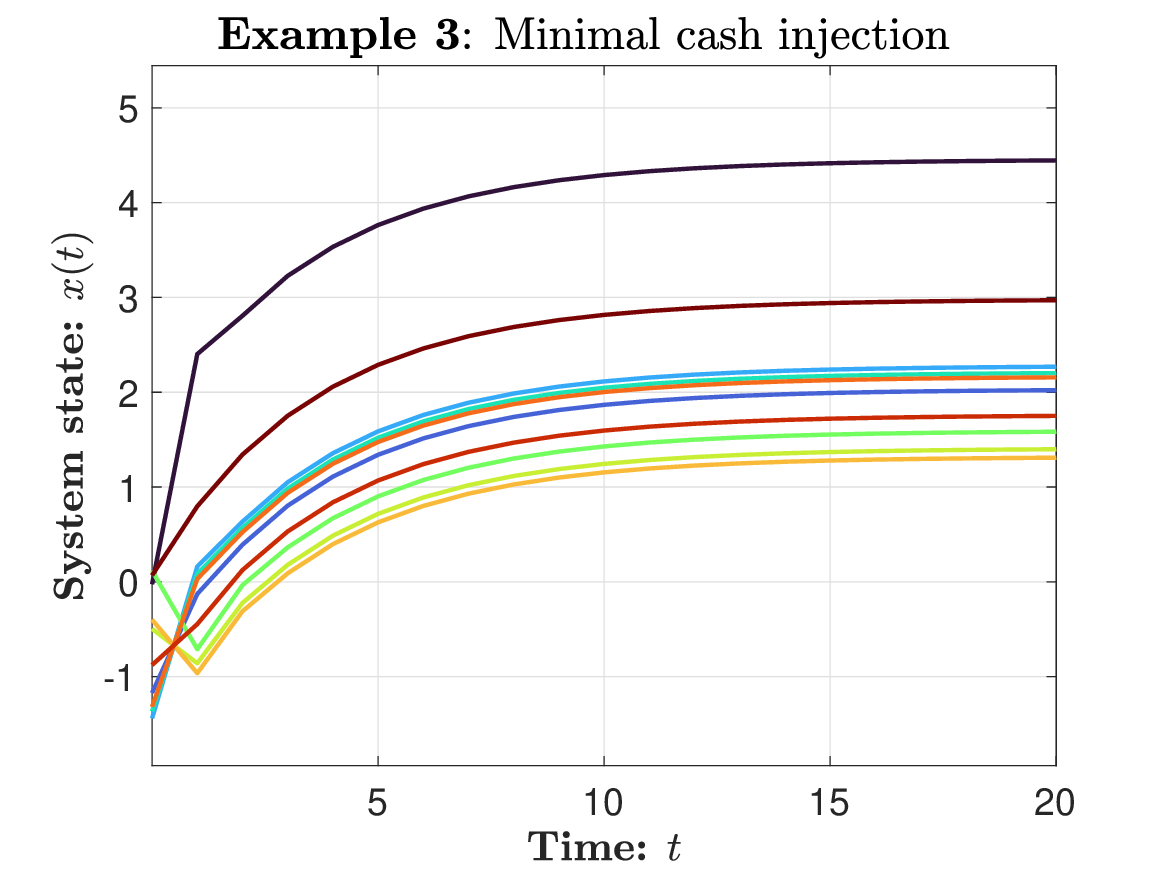}
    \caption{Starting from an initial condition where some of the companies have failed, we find the minimal cash investment that over time brings the system to the maximal positive invariant region $\mathcal M^+$.}
    \label{fig:cash}
\end{figure}
\end{example}

\section{Conclusions}\label{sec:conc}
In this paper, we have investigated the propagation of failures in the global financial system. After introducing the problem and some preliminary results, we have characterized the regions of attractions and invariance of each orthant. Then, we have studied periodic solutions and showed that for periods greater than $2$, such solutions can exist. Finally, we have investigated the problem of finding the minimal cash investment to bring the system in the maximal positive invariant region where all companies are healthy and presented the optimization problem. Future works include the design of the control law to minimize contagion and the study of the corresponding continuous-time system.


\section*{Acknowledgments}
LS has been partly supported by the HUMAT Research Project, financed by the US Army Research Lab, USA, in collaboration with the Alan Turing Institute, UK.
DB has been supported by the SMiLES Research Project, part of the Research Programme Sustainable Living Labs, which is co-financed by the Dutch Research Council (NWO), the Ministry of Infrastructure and Water Management, and the Dutch Institute for Advanced Logistics (TKI Dinalog) under grant 439.18.459. FB and PC have been supported by the European Union  - NextGenerationEU - PNRR M4.C2.1.1 – PRIN 2022 - 2022LP77J4$\_$002 – CUP G53D23000400006 ``Proliferation, resistance and infection dynamics in epidemics (PRIDE)''.



\end{document}